\documentstyle[11pt,newpasp,twoside,epsf]{article}
\def\edcomment#1{\iffalse\marginpar{\raggedright\sl#1\/}\else\relax\fi}
\reversemarginpar

\begin{document}
\title{Evolution of Multipolar Magnetic Fields in Isolated 
Neutron Stars and its effect on Pulsar Radio Emission}
\author{D. Mitra$^1$, S.Konar$^2$, D. Bhattacharya$^1$, A. V. Hoensbroech$^3$,\\
J. H. Seiradakis$^4$ and R. Wielebinski$^3$}
\affil{$^1$Raman Research Institute, Bangalore 560 080, India\\
$^2$Inter University Center for Astronomy and Astrophysics, Pune, India\\
$^3$Max-Planck-Institut f\"{u}r Radioastronomie, Auf dem H\"{u}gel 69, Germany\\
$^4$University of Thessaloniki, Department of Physics, Laboratory 
of Astronomy, GR-54006 Thessaloniki, Greece}

\begin{abstract}
The evolution of the multipolar structure of the magnetic field of isolated 
neutron stars is studied assuming the currents to be confined to the crust. 
Lower orders ($\le 25$) of multipole are seen to evolve in a manner similar to 
the dipole suggesting little or no evolution of the expected pulse shape. We 
also study the multifrequency polarization position angle traverse of PSR B0329+54 
and find a significant frequency dependence above 2.7 GHz. We interpret this as an 
evidence of strong multipolar magnetic field present in the radio emission region. 
\end{abstract}
\vspace{-0.4 in}
\section{Introduction}

Strong multipolar components of the magnetic field have long been thought to play 
an important role in the radio emission from pulsars. Multipolar 
components present even in a small degree can significantly change the 
radius of curvature of the field lines near the polar cap and determine 
the illumination of the pulsar beam. This in 
turn might be responsible for the observed complexity in pulse profiles. 
Significant evolution in the structure of the magnetic field may therefore 
lead to simplification 
of pulse profile with age. We explore the ohmic evolution of multipolar magnetic 
fields in isolated neutron stars assuming the flux to be confined within the crust. 

Multipolar magnetic fields would also cause the polarisation position angle 
(PPA) traverse to depart from the well known rotating vector model of 
Radhakrishnan and Cooke (1969) (RC). As a result of radius-to-frequency mapping 
(RFM) radio emission at higher frequencies originate closer to the stellar 
surface where multipole fields are stronger.  This would lead to a frequency 
dependence of the PPA traverse, departures from the RC model being stronger
at higher frequencies.  We monitor the PPA of PSR B0329+54 at several 
frequencies to investigate this effect. 
\begin{figure}
\plottwo{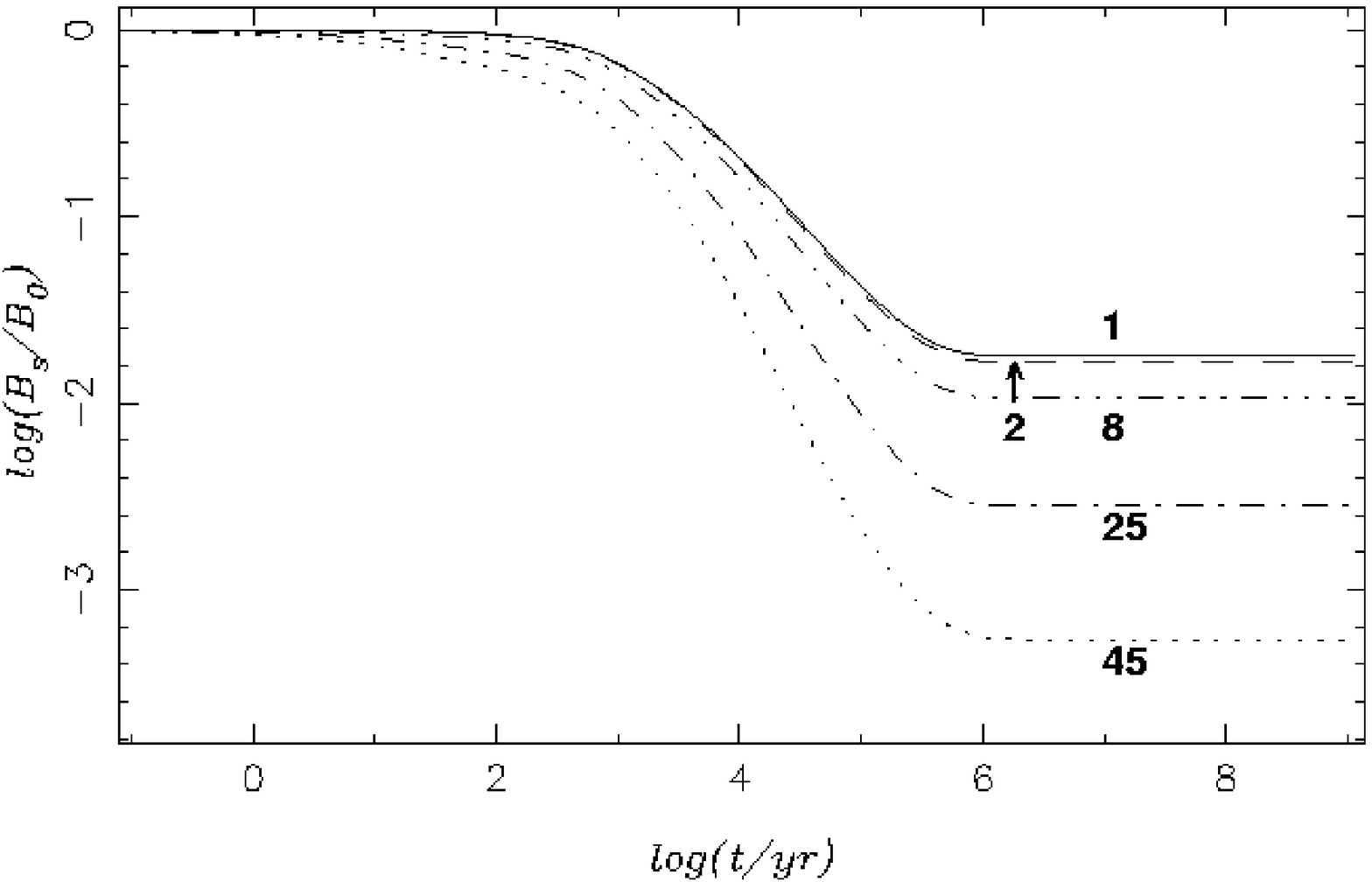}{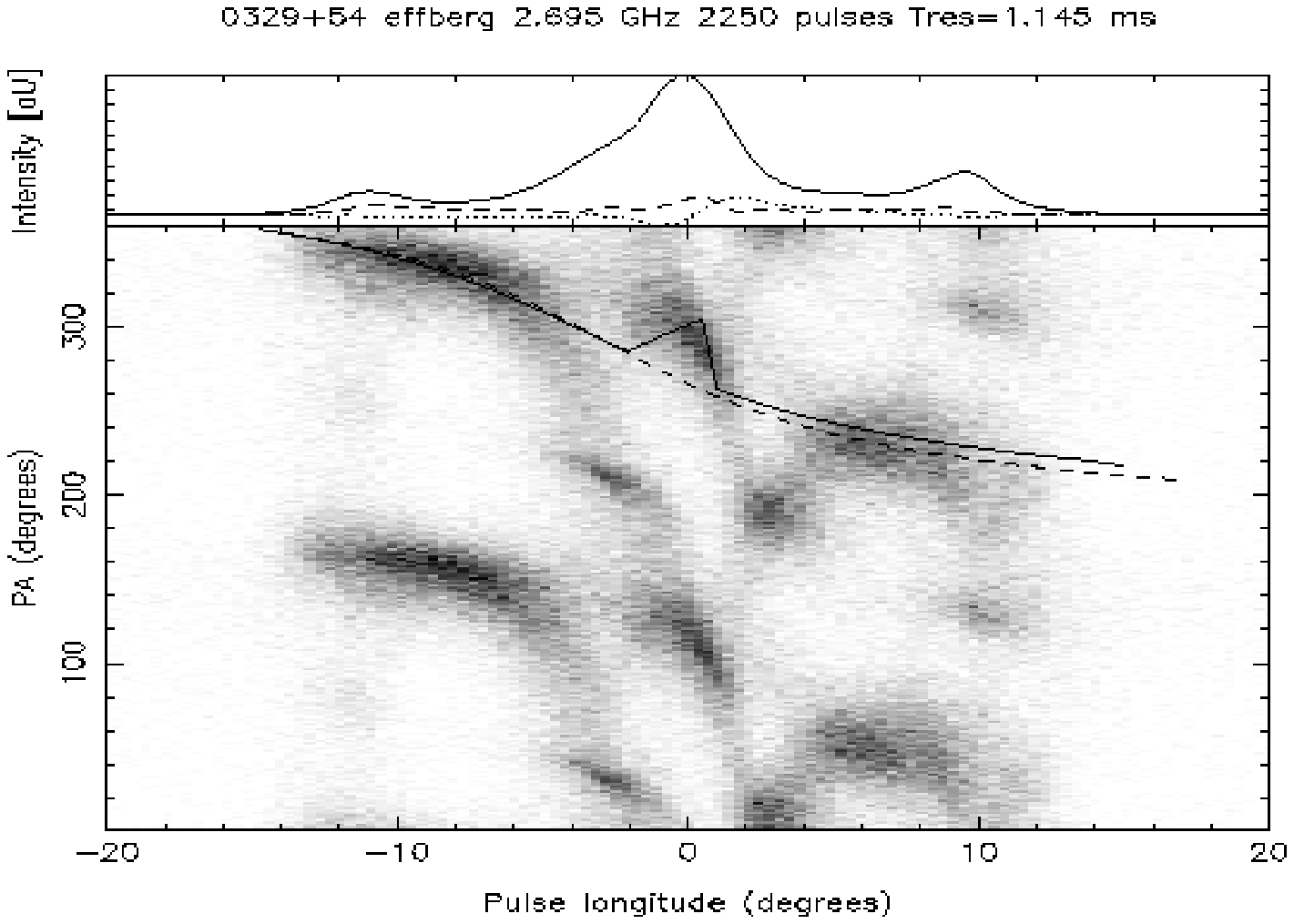}
\caption[]{The figure on the left shows the evolution of various multipole 
orders due to ohmic diffusion. The curves are labeled with multipole orders. 
The figure on the right shows a grey scale representation of the PPA of 
PSR B0329+54 at 2.7 GHz (we have used archival data obtained using the 
Efflesberg Radio Telescope). The dotted line denotes the dipole RC curve as 
obtained at lower frequencies. The continuous line models the PPA traverse 
including a quadrupole component.}
\label{fig1}
\end{figure}
\section{Evolution of Multipolar Magnetic fields}
In order to calculate the evolution of the multipolar magnetic field we solve the 
ohmic diffusion equation numerically. We assume the currents responsible for the 
entire field structure to be confined to the crust of the neutron star. We adopt a 
well motivated choice for the electrical conductivity as a function of 
depth in the neutron star interior and use a standard cooling curve 
to represent the evolution of temperature (for details see Mitra et al. 1999).  

Fig. [1] (left panel) shows the evolution of different multipole orders 
assuming the same initial strength for all orders. It is evident from 
the figure that except for very high orders the reduction in the field 
strength is similar to that of a dipole. 
Hence no significant evolution of the pulse shape is expected due 
to the evolution of the multipolar structure of the field.

\section{Polarization position angle of PSR B0329+54}
We investigate the multifrequency behaviour of the polarization position 
angle of PSR B0329+54 at a number of frequencies over a wide 
range (from 408 MHz to 4.8 GHz). The PPA curves at 408 MHz and 1.4 GHz 
are in agreement with the dipolar RC model but at 2.7 GHz and 4.8 GHz 
the PPA appears to have developed a kinky feature (Fig. [1] 
right panel) which is absent at lower frequencies. Assuming RFM, the observed 
kinkiness might be due to the presence of nondipolar magnetic fields 
in the emission region. The continuous line as shown in Fig. [1] is a 
model for the nondipolar magnetic field structure (a combination of a 
dipole and a quadrupole located at the center of the star) which provides 
a good fit to the observed PPA (Mitra 1999).

\end{document}